\documentclass[a4paper,12pt]{article}
\usepackage[utf8]{inputenc}
\usepackage{cancel}
\usepackage{ulem}
\usepackage{amsfonts}
\usepackage{amssymb}
\usepackage{graphicx}
\usepackage{amsmath,bm}
\usepackage{enumerate}
\usepackage{mathtools}
\usepackage{tikz} \usetikzlibrary{calc}
\usepackage[countmax]{subfloat}
\usepackage{xcolor}
\usepackage{float}
\usepackage{color}
\usepackage{here}
\usepackage{cite}
\usepackage{subcaption}
\usepackage{mwe}

\usepackage{mathrsfs}
\usepackage{float,epsfig}
\usepackage{dcolumn}% Align table columns on decimal point
\usepackage{pgfplots}
\usepackage{graphicx}% Include figure files
\usepackage{bm}% bold math
\usepackage{amsmath,amssymb,amsthm}
\usepackage[colorlinks=true,linkcolor=blue,citecolor=red]{hyperref}
\usepackage{mathtools}

\definecolor{lime}{HTML}{A6CE39}
\newcommand{\orcidicon}{%
	\begin{tikzpicture}
	\draw[lime, fill=lime] (0,0)
	circle [radius=0.16]
	node[white] {{\fontfamily{qag}\selectfont \tiny ID}};
	\draw[white, fill=white] (-0.0625,0.095)
	circle [radius=0.007];
	\end{tikzpicture}   \hspace{-2mm}
}
\newcommand\orcidFaical{{\href{https://orcid.org/0000-0002-2977-0821}{\orcidicon}}}
\title{\bf Reformulation of Classical Thermodynamics from Information Theory}

\author{
	F. Barzi\orcidFaical\!\!$^{1,2}$\thanks{faical.barzi@edu.uiz.ac.ma (corresponding author)} ,   
    K. Fethi\!\!   $\;^{2}$\thanks{fethikaoutar@gmail.com}\\
{\small $^{1}$ LPTHE, Physics Department, Faculty of Sciences, Ibnou Zohr University, Agadir, Morocco. }\\
{\small $^{2}$CRMEF, Regional Center for Education and Training Professions Marrakesh, Morocco.}
}
\date{\today}
\begin{document} 
	\maketitle
\begin{abstract}
		{\noindent In this study, we present a reformulation of classical equilibrium thermodynamics by replacing the obscure and ambiguous concept of entropy with the clear and intuitive concept of information stored in a thermodynamic system. Specifically, we rewrite the laws of thermodynamics in the mathematical terminology borrowed from information theory with an emphasis on information instead of entropy and on binary logarithm instead of natural one. We also define a modified ideal gas constant denoted as $\mathcal{R}$ that quantifies the energy cost of storing or retrieving a mole of information. Moreover an application to the ideal gas is carried and new insight on the evolution of the system is acquired. This new formulation might serve as a basis to teaching thermodynamics, where one deals with the concept of energy and information both stored in a thermodynamic system and avoids the lack of cognitive representations inherent in the concept of entropy.}
\end{abstract}
	%\addcontentsline{toc}{section}{\nameref{appendix}}
	\tableofcontents
	
	%\newpage
	%\newpage

	%	 \tableofcontents

\section{Introduction}
\paragraph{}In classical thermodynamics, entropy is a key idea that is frequently connected to the degree of disorder in a system and the directionality of spontaneous processes. However, because of its various application contexts, the concept of entropy is packed with ambiguity\cite{ben2008c,ben2008cc,carnap2021c,müller2007c}. 
% This inherent vagueness of the concept of entropy is a result of the diverse interpretations that surround it.
Entropy poses a difficult challenge to our comprehension of physical systems because of its definitions, associations with disorder, connection to irreversibility, and numerous applications in various scientific fields\cite{Haynes1980c,Haglund2010c}. This has stimulated deep and often controversial philosophical reflections about the nature of reality, time, and the universe itself, in addition to mystifying the discourse around thermodynamics. Moreover, elucidating the notion of entropy has proven a difficult task and poses numerous obstacles to new learners of thermodynamics as well as to teachers of this important subject. The inability to connect entropy with a familiar day-to-day concept renders its assimilation as a key to understanding thermodynamics an hard goal to achieve. Some recurrent remarks that students give are \textit{what is entropy really?}, \textit{we have a formula to compute entropy but it does not help our understanding much}, \textit{can we have a physical picture of what entropy is?}. When faced with these questions, teachers are incapacitated or may revert to some misleading analogies that raise more questions than they answer. 

\paragraph{}Historically, Rudolf Clausius\cite{clausius1879} was inspired to introduce the concept of entropy in $1865$ through his observations of heat engines and the effectiveness of energy conversion. The development of thermodynamic theory was deeply entrenched in the quest to comprehend and quantify heat transfer, work, and energy transformations\cite{tait1877,wood1889,parker1891,poincaré1892}. The idea was developed further by scientists like Josiah Gibbs, Ludwig Boltzmann, Walther Nernst and Max Planck, which resulted in its dual interpretations in statistical mechanics and classical thermodynamics. This historical trajectory illustrates not just the scientific growth of entropy but also the philosophical problems it raises about energy, work, and the nature of physical systems. Boltzmann in particular, promoted an interpretation of entropy in terms of macrostates and their associated microstates. His famous relation,
\begin{equation}\label{eq0}
    S=k_B\log(\Omega),
\end{equation}
constituted a  breakthrough in our understanding of irreversible processes and the systems' evolution. However, As taught today, classical thermodynamics still don't fully incorporate the implications of his fundamental insight. Most introductory textbooks\cite{whitman2019,Turns2020,saggion2019,Sagawa2009,Nolting2017,lewis2020} on classical thermodynamics only invoke Boltzmann interpretation as a side remark or as a possible alternative to the main old-fashioned disorder-based one. The association of entropy with disorder is among the most widely held interpretations. The tendency of systems to change from ordered to disordered states as they approach thermodynamic equilibrium raises entropy. This formulation seems reasonable at first, but it becomes problematic when one thinks about systems in which entropy increases as order increases. For instance, when water freezes to form ice its entropy decreases, yet the entropy/disorder of the surrounding environment increases due to the release of heat during the freezing process, demonstrating that local decreases in entropy/disorder can occur at the expense of greater increases elsewhere. The intricate connection between disorder and entropy creates uncertainty about the actual nature of entropy in diverse physical contexts. By disregarding the complex relationships that exist between a system and its surroundings, the interpretation of entropy as a measure of disorder can lead one to believe that all spontaneous processes inevitably result in increased disorder. This uncertainty can confound our understanding of events such as phase transitions, chemical reactions, and biological processes.

% \paragraph{}The crucial role that entropy plays in the irreversibility of natural processes is another important feature of this concept. As the second law of thermodynamics states, processes are assumed to evolve toward states of higher entropy/disorder, that is, the total entropy of an isolated system can never decrease. This raises the issue, though,\textit{ what exactly constitutes an isolated system?} Since most systems interact with their environment, it is rare for systems to be completely isolated from their surroundings, which allows for local entropy to decrease even as total entropy rises. Taking the diffusion process as an illustration, the entropy of the system increases when an ink drop is submerged in water because the ink molecules disperse. However, if we observe a container divided into two parts, one filled with ink and the other with water, removing the partition would allow the ink to spread, increasing the entropy. But, if we were to somehow reverse this process, it would require energy input and a decrease in the entropy of the system. Such scenarios highlight the duality of entropy as both a measure of disorder and a guide for the direction of thermodynamic processes, thereby complicating the application of the second law.

\paragraph{} In information theory, entropy measures the \textit{information content of a system}, a concept introduced by Claude Shannon in the mid-$20^{th}$ century\cite{Shannon1948}. The subsequent work of Khinchin\cite{Khinchin1957} provided mathematical rigor to the theory. Since then, information theory has become the best framework to interpret entropy and its manifestations in different physical and non-physical situations. Nevertheless, classical thermodynamics remained in some extent outside this ground-breaking progress. As seen in all textbooks on the subject, the old formulation still retains the conceptual idea of entropy inherited from the industrial revolution with all its shortcomings.  In this study we propose a reformulation of classical thermodynamics where the revolutionary concept of information supersedes the ambiguous concept of entropy. Such reformulation is required in the information age to give physics teachers and their prospective students the correct viewpoint to address information-related challenges.

\section{Number of moles of information stored in a system}
At thermodynamic equilibrium the entropy of a thermodynamic system is given by Boltzmann equation, Eq.\eqref{eq0}, 
% \begin{equation}\label{eq1}
%    \displaystyle S=k_B\log(\Omega)
% \end{equation}
where $\Omega$ is the number of microstates compatible with the thermodynamic equilibrium macrostate and $k_B$ is the Boltzmann constant. From information theory, Shannon showed that Eq.\eqref{eq0} is a particular case of a more general equation for entropy,

\begin{equation}
    S=-k_B\underset{i}{\sum}p_i\log(p_i)
\end{equation}
where $p_i$ is the probability of occurrence of the $i$-$th$ microstate of the thermodynamic system. In this formula, entropy is a measure of the information content of the system. To initiate the reformulation, we define a more transparent quantity by considering the number of bits $\alpha$ needed to associate a binary encoding to each microstate such as,

\begin{equation}\label{eq2}
   \displaystyle \Omega=2^\alpha
\end{equation}

Concretely, a binary encoding is a sequence of zeros and ones that uniquely labels each microstate. for example, $\alpha=2$ means that the system has $2^2=4$ micro-states with binary encoding $00,\,01,\,10,\,11$. Then the entropy can be written as,
\begin{align}
    S&=k_B\log(2^\alpha)\\
    &=k_B\log(2)\,\log_2(2^\alpha)\\
    &=k_B\log(2)\, \alpha\\
    &=\bm{\mathcal{R}\, n_b},\label{eq8}
\end{align}
where we define the quantity,
\begin{equation}
    \displaystyle n_b=\frac{\alpha}{N_A},
\end{equation} 

as \textit{the number of moles of information\footnote{We might also use interchangeably \textit{the number of bits of information}.}} measured in \textbf{\textit{mole-bit}} (\textbf{\textit{molb}}),
\begin{center}
    \textit{$1$ molb = $1$ mole of bits of information}
\end{center} 
and
\begin{align}
    \mathcal{R}=k_B N_A\log(2)\simeq5.76315 \;J\, K^{-1}\,molb^{-1},
\end{align}

is the \textit{modified constant of ideal gases}. Eq.\eqref{eq8} sets up the starting point for a more intuitive reformulation of classical thermodynamics based on the concept of the number of moles of information.
It is well known that for an isolated thermodynamic system evolving towards thermodynamics equilibrium, the number of microstates increases, and so, the number of bits to encode these new microstates, that is, $n_b$ must increase. We propose to replace the obscure concept of entropy $S$ by the more transparent concept of \textit{number of mole of information stored in the thermodynamic system $n_b$}.

\section{Reformulation of the laws of classical thermodynamics}
\paragraph{}In this section we propose a systematic reformulation of the laws of thermodynamics to be compatible with insights from information theory. we start with the second law. The zeroth law which postulate the notion thermal equilibrium and the concept of temperature is supposed unchanged.
\subsection{Second law of thermodynamics}
\paragraph{}The second law of thermodynamics is reformulated such as, 
\begin{center}
    \textbf{\textit{for an isolated system, the number of moles of information stored in the system must increase, $\bm{\Delta n_b\geq0}$}}
\end{center}
In the case of a closed system exchanging heat $Q$ with its surrounding, we have,\\

    \textbf{\textit{ During a thermodynamic process, the variation of the number of moles of information $\bm{\Delta n_b}$, measured in molb, due to the heat exchange $Q$, measured in Joules, is given by,}}
\begin{equation}
    \bm{\Delta n_b=\frac{Q}{\mathcal{R}\,T_{int}}+n_{b}^i}
\end{equation}
\textbf{\textit{where $\bm{T_{int}}$ is the temperature of the exchange interface, measured in Kelvin, and $\bm{n_{b}^i\geq0}$ is the an intrinsic increase of the number of moles of information, measured in molb. For a reversible thermodynamic process, $\bm{n_{b}^i=0}$.}}

\paragraph{}In this reformulation, as the system exchanges heat with its surrounding, the number of bits of information necessary to encode the microstates changes accordingly, such that, if the system gains some heat $Q>0$, the number of bits increases and vice versa. The quantity $\mathcal{R}T_{int}$, acquires the important meaning of \textit{energy cost per mole of information}, That is, for each mole of information gained or lost by the system, it must exchange a heat equal to $\mathcal{R}T_{int}$. Additionally, the \textit{entropy creation} which usually bewilders students find here a more satisfying picture as an intrinsic increase in the amount of information stored in the system due to internal dynamics, like chemical or nuclear reactions that create new microstates. During its evolution to thermodynamic equilibrium, an isolated system evolve from one microstate to another. At a given time $t$, it has $2^{n_b}$ available microstates to move to, however at $t+\delta t$ it will have $2^{n_b+\delta n_b^i}$ possible microstates. The information content of the isolated system increases steadily to encode in binary these new microstates.

\paragraph{}According to Landauer's principle\cite{Landauer1961,Bennett2003}, the minimal energy cost of erasing information is $k_B \,\log(2) T_{int}$, or the increase in entropy of the environment caused by erasing a single bit of information. This principle shows that information is not just abstract but has physical consequences, which emphasizes the thermodynamic significance of information processing.
Information theory and thermodynamics are directly related to Landauer's principle, which has been experimentally confirmed in a number of settings \cite{Bennett1982,Sagawa2009,Faist2015}. This link casts doubt on the conventional understanding of computation as an abstract, non-physical activity by highlighting the cost of computation and the physicality of information. This shows the central role of information in the interpretation of any heat exchange. Specifically, the heat exchange is a manifestation of the gain or loss of information by the system measured in number of bits (or molb).  The entropy in this regards is an redundant concept inherited from the industrial revolution and is prone to misleading and difficult interpretation. 

\paragraph{}While teaching thermodynamics, it is hard to give a clear meaning to entropy\cite{Baierlein1994c} and often it appears as a miraculous quantity that was discovered by the forefathers of thermodynamics and happens to indicate the evolution of systems. Information on the contrary is easy to convey and comprehend allowing the teacher and the student to focus more on the physical implications of the laws not on the laws themselves. One can also use another terminology when dealing with bits of information, that is, a mole of information (a molb) is \textit{retrieved from} or \textit{stored in} the system requires a heat exchange \textit{lost} or \textit{gained} by the system, respectively.
\subsection{First law of thermodynamics}
\paragraph{}The first law can also be rewritten accordingly. The variation of total energy $\Delta E$ of the closed thermodynamic system due to work and heat exchanges is given by,

\begin{align}
 \Delta E&=Q+W\\
 &=\mathcal{R}\,T_{int}\Delta n_b-p_{int}\Delta V
\end{align}

It is important to stress again the meaning of information in this formalism. We have replaced the obscure concept of entropy in order to gain more insight into the time evolution of thermodynamic system. Therefore, one must be clear on the significance of the information in the present context. The number $n_b$ measures the quantity of \textit{information stored in the system} and\textit{ not information about the system}, these two kind of information should be sharply distinguished. Information about the system is the knowledge about its current microstate. This information decreases as an isolated system evolve towards thermodynamic equilibrium because more and more microstates are created and the observer knows less and less about the true microstate. However, the quantity of information in the system increases due to the increase of the number of microstates and the need to encode them require more bits of information. Once this distinction is made there is no more ambiguity and one has a clear cut concept to work with, that is, \textit{information stored in the system}.

\paragraph{}As a mental representation, it is easier for the student to imagine the thermodynamic system having total energy and information as state quantities that are modified by the exchange of work and heat. Energy and Information are two concepts that the student can relate to and has some daily awareness of. Entropy on the other hand appears totally foreign and hinges on philosophical discourse that neither the teacher can satisfactory explain nor the student can be expected to comprehend. Moreover entropy is full with mysticism from media sensation and popular culture mystification; hence, before a student even begins to study thermodynamics, a psychological barrier may already exist.

\paragraph{}From Eq.\eqref{eq2}, it is clear that the number of bits $\alpha$ is not in general an integer. To illustrate how the variation of the number of bits changes during thermodynamic evolution, we suppose that $\alpha$ was initially equal to $2$, that is we need $2$ bits of information to encode the $4$ microstates $00,\,01,\,10,\,11$ of the system. Once the number of these microstates reaches $5$, we need $3$ bits to encode all microstates. Explicitly, they acquire the encoding $000,\,001,\,010,\,011,\,100$. We see that fundamentally the number of bits changes by $1$ to accommodate the new microstate even if some encoding are not attributed to any microstate. Therefore, it is natural to  define \textit{the actual number of bits} denoted as $N$,
\begin{equation}
   \displaystyle N=\lceil\alpha\rceil
\end{equation}
here $\lceil.\rceil$ is the \textit{ceil function} and $N$ is the smallest integer larger than $\alpha$. In the above example, $\alpha\simeq2.321$ at $5$ microstates, since $5\simeq2^{2.321}$, thus $N=\lceil2.321\rceil=3$. Consequently, the actual number of bits increases by steps of integer numbers.
\subsection{Third law of thermodynamics}
\paragraph{}The third law of thermodynamics which provides an absolute reference for entropy can be reformulated such that it defines an absolute reference for the information stored in a thermodynamic system,

\begin{center}
      \textbf{\textit{ As temperature falls to zero, the number of moles of information stored in any pure, perfect and homogeneous crystal tends to a universal constant.}}
    \end{center}
\begin{equation}
    \bm{T\longrightarrow0\implies n_b\longrightarrow n_0}
\end{equation}

In this form, the third law asserts that one can not extract all the information stored in a pure crystal. At least one bit of information is needed to encode the unique microstate by the encoding $0$ or $1$. Implying that the actual number of bits at $T=0$ is $1$, $N=1$. Otherwise, a degeneracy of the ground state would require more bits of information. It is worth noting that as temperature approaches absolute zero, the energy cost per bit of information goes to zero and therefore it costs virtually nothing to extract a bit of information from the thermodynamic system. In other words, at absolute zero, retrieving/storing a bit of information does not involve a heat exchange. Thus operating in this regime from an information processing point of view is the most profitable.

\paragraph{}Restatements of other versions of the third law supply additional insight into its meaning in terms of information stored in the system. The Nernst's formulation\cite{Nernst1926} deals with thermodynamic processes at a fixed and low temperature for condensed systems such as liquids and solids,
\begin{center}
    \textbf{\textit{The change in the number of moles of information stored in any condensed system undergoing a reversible isothermal process approaches zero as the temperature at which it is carried approaches absolute zero.}}

\vspace{0.5cm}
    Or equivalently,
\vspace{0.5cm}

\textit{\textbf{At absolute zero, the change in moles of information stored becomes independent of the process path.}}
\end{center}

There is also the unattainability formulation due also to Nernst,
\begin{center}
    \textit{\textbf{It is impossible for any process, no matter how idealized, to reduce the number of moles of information stored in a system to its absolute-zero value in a finite number of operations}}
\end{center}
 Another simpler formulation equivalent to Planck's version and often attributed to Einstein states,

 \begin{center}
     \textit{\textbf{The number of moles of information stored in any substance approaches a finite value as the temperature approaches absolute zero.}}
 \end{center}

\section{Application to the ideal gas}
In this section we apply the new formulation to the thermodynamics of the ideal gas and compute the number of mole-bits $n_b$. From the first law Eq.\eqref{eq2}, we have for a infinitesimal process of $n$ moles of an ideal gas with temperature $T$ and pressure $p$,
\begin{equation}
dE=\mathcal{R}\,T \,dn_b-p\,dV.
\end{equation}
We suppose the ideal gas is at rest with respect to the laboratory frame of reference. Thus, only the internal energy of the system changes,
\begin{equation}
dU=nc_vdT=\mathcal{R}\,T \,dn_b-p\,dV,
\end{equation}

where $c_v$ is the molar heat capacity at constant volume. We have for the number of mole-bits,
\begin{equation}
dn_b =n\,c_v\frac{dT}{\mathcal{R}\,T} + p\,\frac{dV}{\mathcal{R}\,T}
\end{equation}

Using the equation of state of the ideal gas, $p\,V=n\,R\,T$, we get,
\begin{equation}
dn_b =\frac{n\,c_v}{\mathcal{R}}\frac{dT}{T} + \frac{n\,R\,}{\mathcal{R}\,}\frac{dV}{V}
\end{equation}

\begin{figure}[ht!]
    \centering
    \includegraphics[scale=0.75]{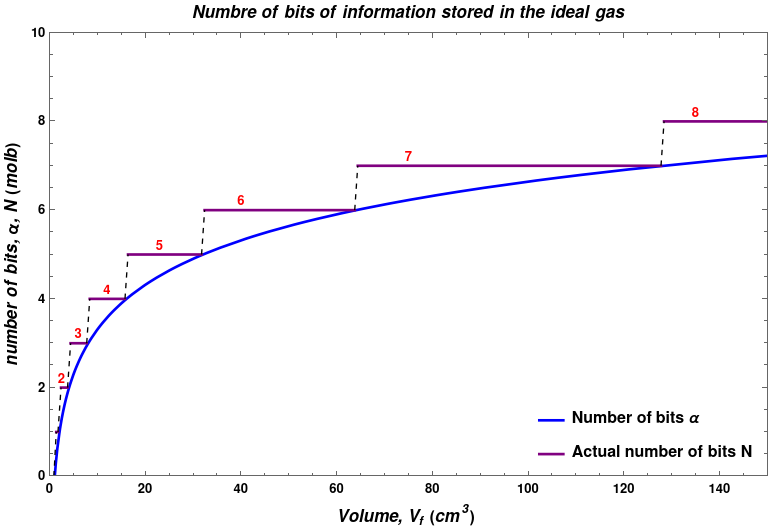}
    \caption{\footnotesize{\it Evolution of number of bits stored in the ideal gas with the volume. The blue line is the number of bits of information stored in the gas, while the purple piece-wise line is the actual number of bits to encode the microstates of the gas.}}
    \label{fig:1}
\end{figure}

After an integration from an initial state $(i)$ to a final state $(f)$, we obtain the following expression for $n_b$,

\begin{align}
 \Delta n_b&=\frac{n\,c_v}{\mathcal{R}}\log\left(\frac{T_f}{T_i}\right) + \frac{n\,R\,}{\mathcal{R}\,}\log\left(\frac{V_f}{V_i}\right)\\
 &=\frac{n\,c_v}{R}\log_2\left(\frac{T_f}{T_i}\right) + n\log_2\left(\frac{V_f}{V_i}\right)\label{eq3a}
\end{align}
where we use the binary logarithm in Eq.\eqref{eq3a}. Let's suppose that the ideal gas undergoes \textit{\textbf{a mechanically reversible and isothermal expansion at temperature $\bm{T_0}$}}. Then the change in number of bits is,
\begin{equation}
    \Delta n_b= n\log_2\left(\frac{V_f}{V_i}\right)\label{eq4a}
\end{equation}
It is made manifest that the number of bits of information stored in the system is related to the number of constituents, that is, the more constituents a system has the more information can be stored in it. this is a much more simpler to interpret and visualize for a student than the concept of entropy. This also shows that \textit{mole-bit} and \textit{mole} are the same unit, however, it is important to keep a distinction between \textit{bits of information} and \textit{constituents of matter}, suffices to stress that they are nonetheless intimately connected. To graphically depict the evolution of the actual number of bits $N$, we suppose that $n=1\, mol$ and $V_i=1$ $cm^3$. Then, Eq.\eqref{eq4a} becomes,
\begin{equation}
      \Delta N= N_A\lceil\log_2\left(V_f\right)\rceil\label{eq4}
\end{equation}
Fig.\ref{fig:1} shows the variation of the number of bits $\alpha$ and the actual number of bits $N$ with the volume of the gas. We recall that $2^\alpha$ is the number of microstates of the system and $N$ is the number of bits needed to unambiguously encode all these microstates in binary. 

\paragraph{}During its  expansion, the ideal gas  exchanges heat and work with its surrounding such as,
\begin{equation}
     dU=0 \implies \delta Q=-\delta W=\mathcal{R}T_0 \left(dn_b-\delta n_b^i\right)
\end{equation}

On the other hand work exchange is given as,
\begin{equation}
    \delta W=-RT_0\frac{dV}{V}\implies W=-RT_0\log\left(\frac{V_f}{V_i}\right)=-\mathcal{R}T_0\log_2\left(\frac{V_f}{V_i}\right)
\end{equation}
We notice that work exchange is expressed in terms of logarithm in base $2$ in the same way we used to in terms of natural logarithm with the replacement $R\longrightarrow\mathcal{R}$. Now the heat exchange is,
\begin{equation}
    Q=\mathcal{R}T_0\log_2\left(\frac{V_f}{V_i}\right)=\mathcal{R}T_0\left(\Delta n_b-n_b^i\right)
\end{equation}

Using Eq.\eqref{eq4a} for $n=1\,mole$ we see readily that $n_b^i=0$. That is, \textit{an ideal gas has no intrinsic increase of the number of bits when submitted to an mechanically reversible and isothermal expansion}. Now suppose the ideal gas undergoes a \textbf{\textit{reversible and adiabatic expansion}}. According to the reformulated second law, the number of bits of information does not change, $\Delta n_b=0$, From Eq.\eqref{eq3a} we have,
\begin{align}
    0&=\frac{\,c_v}{R}\log_2\left(\frac{T_f}{T_i}\right) + \log_2\left(\frac{V_f}{V_i}\right)\label{eq3}\\
    &\implies T\,V^{\gamma-1}=cst
\end{align}

Or equivalently, 
\begin{equation}
    p\,V^{\gamma}=cst \quad \text{and} \quad T^\gamma\,p^{1-\gamma}=cst
\end{equation}

In textbooks such process is called an \textit{isentropic process}. In the present reformulation it is an \textit{iso-information process} since the information stored in the system does not change.

\section{Conclusion}
\paragraph{}In this paper, we proposed a reformulation of classical thermodynamics compatible with the insights gained from information theory. All three laws of thermodynamics find a new intuitive restatements in terms of the information stored in a thermodynamic system. Arguably such a recasting of thermodynamics in the terminology of information theory is much more appealing to students and physics teachers, mainly due to the plainness and clarity presented by the concept of information in contrast to the ambiguity inherent in entropy.

\paragraph{}We proceeded to apply the new formulation to the ideal gas system and showed that one readily gains more insight by invoking the information stored in the gas by building simple and accurate mental representations  of the evolution of the system. The evolution of a system is driven by a change in its information content, a more tangible concept than entropy, either by retrieval or storage of bits of information to encode uniquely its microstates. The manifestation of such processes in the form of heat exchange should not hide the central role played by information. Teaching classical thermodynamics in this form might greatly simplify the task on both sides of the learning process. One then deals with two state functions, \textit{total energy} and \textit{total information}, both stored in the system. The heat and work exchanges are the expressions of their variations during a thermodynamic process.
% \newpage

% \appendix
% \numberwithin{equation}{section}
% \newpage

\bibliographystyle{unsrt}
\bibliography{Reformulation} 
\end{document}